\documentclass[12pt,thmsa]{article}
 \usepackage{amsmath}

 \usepackage{amssymb}

\input{tcilatex}

\begin{document}

\baselineskip 17pt

\noindent \centerline {{\huge {\bf Length contraction puzzle
solved?}}}

\vspace {13mm}

\centerline {\bf Dragan V Red\v zi\' c}

\noindent {\small Faculty of Physics, University of Belgrade, PO Box
368, 11001 Beograd, Serbia, Yugoslavia}\\ E-mail address:
redzic@ff.bg.ac.rs

\vspace {7mm}

\noindent {\bf Abstract}

\noindent It is argued that the `perspectival change' of a physical
object in special relativity may be given a natural dynamical
explanation in terms of a change in the object under the action of
certain forces in a rest properties-preserving way.

\bigskip

\noindent {\bf 1. Introduction}

\noindent In a recent discussion about the meaning of Lorentz
contraction, Franklin [1] stressed that in special relativity (SR)
there is no change in the object, it is only the reference frame
that is changed. In another recent paper, Miller [2] expressed
essentially the same point arguing that `when an observer changes
frames (or when we compare the results of observers in different
frames) there are no dynamical effects in the physical object being
observed ... The different inertial observers have no dynamical
explanation [of the differences among their observations] in terms
of a change in the object, nor do they require one.'

The statement that there is no change in the object in SR is a
commonplace in the literature. (The length of a uniformly moving rod
is reduced as compared with its length in the rod's rest frame, `but
of course nothing at all has happened to the rod itself' [3].) The
statement appears to be truism in the sense that apparently no
action was exerted upon the object by a mere different choice of
reference frame. The differences among the observations of the
different observers are interpreted as natural consequences of a
{\it change in perspective} and thus outside the scope of the
concepts of cause and effect.

Recall briefly where does the `perspectival' interpretation of SR
come from. SR is a `principle' theory, based on the relativity
postulate and the light postulate. From the two postulates it is
deduced that the space-time coordinates of two inertial frames in
relative motion are related by a Lorentz transformation and,
consequently, that any mathematical relationship which is a
candidate to be a law of physics must be Lorentz-covariant. Stress
that in the principle approach to SR it is taken (more or less
tacitly) that measuring instruments (metre sticks and clocks) in any
inertial frame can be constructed {\it in the same way} `from
scratch'.\footnote [1] {It should be noted that the author of SR
originally had not seen provision of measuring instruments in
various frames in this way [4], but it seems that later he fully
embraced this point of view.} Since the above argument is
`independent of any special assumptions about the constitution of
matter' and no explicit mention of forces is found in it, it is
inferred that the Lorentz transformation, as well as its simple
consequences length contraction and time dilation, belong to
geometry, i. e. kinematics of SR. This attitude was concreted when
the Minkowski spacetime with its geometry is taken to be the
fundamental entity that determines the kinematics in all inertial
frames.\footnote [2] {That includes the kinematic relations between
different inertial frames, the Lorentz transformations corresponding
to rotation in the Minkowski spacetime. No dynamical explanation of
these relations is needed as these are encoded in the Minkowski {\it
geometry} that encapsulates the theory's fundamental {\it
principles}.} Thus a change of reference frame (rotation in the
Minkowski spacetime) is naturally interpreted as a change in
perspective, the object being perfectly passive in the process,
without any forces acting on it.

Now, the notorious problem with SR is that `very many people
understand nothing in the beginning but become accustomed to it in
the end'. Einstein's original definition of time [4] and the light
postulate, on its own completely benign, in combination with the
relativity postulate always gives rise to the same dramatic effect:
the feeling of loosing all the ground under one's feet, disbelief
and insecurity, and a perennial question if it is possible that
everything could really be so. Even when this new concept of time is
somehow `swallowed' and the student of relativity yielded to his or
her destiny expects new relativistic wonders, the disbelief and
insecurity stay.

It seems that the discomfort that physicists (and laymen) feel about
the contraction of a rod in motion and the slowing down of a clock
in motion is a consequence of the {\it opacity} of the usual
relativistic method of inference. Namely, the features of a certain
physical system (e. g. a specific moving clock) are deduced not from
the structure of the system described in the inertial frame relative
to which the clock moves (`the lab'), but from the Lorentz
transformations that connect the lab frame and the clock's rest
frame. Naturally, a question arises of what is the role of the clock
frame, with all of its Einstein-synchronized clocks (which, while
mutually identical, may be different from the observed clock in
motion). Is the lab frame not quite sufficient? The Lorentz
transformations appear as the Fates whose power over the destiny of
all physical systems (our moving clock included) is indubitable (as
proven by experiments), but quite puzzling. Einstein himself pointed
out this fundamental limitation of `the principle of relativity,
together with the principle of constancy of the velocity of light'
[5].

From its advent until today, various authors argued with various
degrees of sophistication that one of barriers to understanding of
SR is due to the neglect of its dynamical content, which remains
hidden or implicit in a purely principle approach [6-13], [2].
Namely, despite its precision and power (and perhaps just because of
that), the principle approach to SR which apparently excludes
dynamics may give rise to fundamental misconceptions. For example,
the fascinating simplicity and universality of Einstein's original
derivation of length contraction was a kind of red herring: the
derivation of the phenomenon is taken to be its root. Thus length
contraction is interpreted as a kinematical effect whereas in fact
dynamical concepts are indispensable for the right
interpretation.\footnote [3] {The length $L_0'$ of a rod in its rest
frame $S'$, and the length $L_v$ of the rod in the frame $S$
relative to which it moves along its length at the speed $v$, are
related by

$$
L_v = L_0'\sqrt{1 - v^2/c^2}\, . \qquad (A)
$$

\noindent The length $L_v$ of the rod in motion is determined by
{\it equilibrium} of internal forces in $S$ governing its structure;
{\it mutatis mutandis} the same remark applies to $L_0'$. Since
$L_v$ and  $L_0'$ are determined by the forces, the simple
relationship (A) between the two lengths is due to the fact that the
forces must be Lorentz-covariant [7]. (A witty model of a measuring
rod illustrating this is found in [2].) Thus equation (A) is a
consequence of properties of force fields in equilibrium (statics,
and not kinematics, as Fe\u{\i}nberg [10] pointed out).

True, one could argue that Lorentz covariance of the force fields is
a consequence of kinematics, i. e. properties of the Minkowski
spacetime. I believe, however, that this would involve reducing
physical concepts such as length to geometry, i. e. reification of
the Minkowski {\it geometry}. }

As a remedy, Bell [9] advocated the use of a constructive
(dynamical) approach in teaching SR (`Lorentzian pedagogy'):
starting from known and conjectured laws of physics in any {\it one}
inertial frame, one can account for all physical phenomena,
including the experience of moving observers. He attempted to
illustrate this programme by considering a simple model of the
hydrogen atom in the framework of Maxwell's theory of
electromagnetism,\footnote [4] {As Fe\u{\i}nberg [10] pointed out,
it is a miracle that Maxwell had written his equations in a
Lorentz-covariant form straightaway, luckily not adding some extra
terms in them. It is clear that Bell's constructive approach,
following the path made by FitzGerald, Larmor, Lorentz and
Poncar\'{e}, was possible due to the happy circumstance that
Maxwell's equations were Lorentz-covariant (even better, they were
Lorentz-covariant when nobody was aware of that).} anticipating from
the outset the relativistic form of Newton's second law.\footnote
[5] {The relativistic equation of motion is indispensable in a
constructive approach to SR. Namely, one can infer properties of a
physical system in motion after accelerating it starting from rest
until reaching a steady state only if physical laws governing its
structure (Maxwell's equations) and the correct relativistic
equation of motion (the Lorentz force equation) are known from the
outset. Then, as Bell outlined, in the long run Lorentz-covariance
of Maxwell's equations follows as the exact mathematical fact which
can be given a natural physical interpretation. Thus Bell's
anticipation of the relativistic equation of motion can not be
considered as a limitation of Bell's approach, contrary to Miller's
statement in [2]. In his recent attempt to derive SR constructively,
Miller [2] avoided the use of the Lorentz force equation. Instead,
he tacitly postulated that Maxwell's equations apply not only in the
original rest frame of a physical object but also in its final rest
frame, cf the argument leading to equation (6) in [2]. Thus Miller
postulated the observations of a moving observer instead of deducing
them, which is hardly a constructive approach to SR. It is rather a
combination of a constructive and a principle approach.}
Unfortunately, even the simple model of the atom is too complex to
be solved analytically, and effects of accelerating the atom can be
painfully recognized only through a numerical solution. Also, anyone
who has ventured to show Lorentz-covariance of Maxwell's equations
and the Lorentz force equation in Einstein's original 1905 way,
without the luxury of tensors in the Minkowski spacetime, knows well
that the task is a true {\it tour de force} [14-16]. Thus Bell's
seminal essay gives only an outline of a constructive approach to
SR. One of the key insights of the paper is the recognition that at
one point of the constructive argument one must postulate
Lorentz-covariance of the complete theory (since the Maxwell-Lorentz
theory provides a very inadequate model of matter). While the
constructive approach must eventually be complemented by the
principle approach, it does feed our lust for meaning: unexpected
properties of rods and clocks in motion do not appear as a dry
consequence of certain abstract mathematical transformations,
achieved from logically entangled postulates, as is the case in
Einstein's approach, but as a natural offspring of earlier physical
ideas.\footnote [6] {Even with an oversimplified model of matter
such as the Maxwell-Lorentz theory of electromagnetism, it is a
revealing small exercise for the student to recognize that using his
or her FitzGerald-Lorentz contracted measuring rods and Larmor
dilated clocks (e. g., those proposed in [2]) a moving observer
would measure that one clock-two way speed of light is again $c$,
the same that was found when the observer, the rods and the clock
were at rest.}

The purpose of the present note is to point out a dynamical aspect
of SR which seems to have been overlooked or perhaps not
sufficiently stressed in the literature. It will be argued that the
differences among observations of observers in different inertial
frames (or when a single observer changes frames) can be interpreted
in a legitimate way in terms of the action of certain forces on the
physical object being observed, regardless of its nature. In other
words, as if something has happened to the object in a change of
reference frame; the so-called `change in perspective' may be
understood as hiding a complex dynamical process. Hopefully, our
treatment could to some extent dispel `the mystical mist' which
surrounds length contraction and time dilation from the advent of
SR.

\medskip

\noindent {\bf 2. Where do the `perspectival' effects come from?}

\noindent Consider two inertial reference frames $S$ and $S'$ in
standard configuration, $S'$ is uniformly moving at speed $v$ along
the common positive $x, x'$-axes, and the $y$- and $z$-axis of $S$
are parallel to the $y'$- and $z'$-axis of $S'$, respectively. As
was noted above, in a purely principle approach to SR, the two
frames are introduced by {\it fiat}, taking that measuring
instruments (metre sticks and clocks) in both frames are constructed
{\it in the same way} `from scratch'. On the other hand, according
to Einstein's original 1905 argument [4], we have initially two
inertial frames in relative rest, each frame being provided with its
own set of measuring instruments, the frames (including instruments)
being identical to one another. Then say the $S'$ frame has been
created by accelerating one of the copy frames (with its set of
instruments), whereas the other copy, its instruments included,
remained at rest (the $S$ frame). Stress that the two ways of
introducing the $S$ and $S'$ frames are equivalent under the proviso
that accelerations were rest properties-preserving.\footnote [7]
{Starting from Einstein [4], various presentations of SR introduce,
often tacitly, the assumption that rest properties of an initially
free connected object in a steady state are preserved under
arbitrary accelerations, if the object is free in the final steady
state (after all transient effects of acceleration have died out).
It seems, however, that construction of SR requires only rest
properties-preserving accelerations [7], [9], [13].}

Consider now a free connected object in an equilibrium internal
state at rest in the $S$ frame, and assume that its rest properties
are known to us. Let the object be accelerated in an arbitrary way
along the $x$-axis in the direction of the increasing $x$ until
reaching steady speed $v$, all with respect to $S$, so that $S'$ is
its new rest frame. Properties of the object in uniform motion
relative to $S$ could be found in the following way, at least in
principle: From known laws governing the structure of the object,
and from known fields of forces accelerating it, using the correct
equations of motion, one could deduce exactly what changes happen in
the object during its acceleration, under the action of external and
internal forces, until reaching the final equilibrium state. In
this, dynamical approach to the problem, all we need are the {\it
true} laws of physics in the $S$ frame and an omnipotent
mathematician. The dynamical approach clearly shows that properties
of the object change until reaching a persistent final state, all
relative to $S$, due to the interplay of external and internal
forces.

In case the object is accelerated in a rest properties-preserving
way, there is another method for finding final steady properties of
the object in uniform motion relative to $S$. As is well known, the
method is provided by SR: starting from known properties of the
object in its rest frame $S'$ one can deduce required properties of
the moving object in $S$ using the laws of transformation of the
relevant physical quantities with respect to the Lorentz
transformation. This principle approach circumvents too cumbersome
(in fact impracticable) calculations appearing in the dynamical
approach.\footnote [8] {The relativity postulate together with the
light postulate plays a role analogous to that of the law of
conservation of energy in mechanics. Namely, certain aspects of a
mechanical problem can be reached in a simple and elegant way
without entering complex dynamical analyses. This point was
discussed in detail in [10].} The price to be paid is a potential
loss of understanding.

In the preceding paragraph a physical object was transferred from
its initial rest frame $S$ to its final rest frame $S'$ through a
rest properties-preserving acceleration. Fe\u{\i}nberg [10] pointed
out, revitalizing Einstein's original 1905 argument, that the same
final state of the object in motion relative to $S$ could be reached
in a different way. Let the object be initially at rest in $S'$ and
let there be another inertial frame also at rest relative to $S'$,
the two frames (including their measuring instruments) being
identical to one another. Now accelerate the copy frame (as well as
its set of measuring instruments)in a rest properties-preserving way
with respect to $S'$ in the direction of the decreasing $x'$ until
reaching the steady speed $v$, without touching the object (which
remains at rest relative to the inertial frame $S'$). Since the copy
frame eventually becomes the frame $S$, we have again the object in
the same uniform motion with respect to $S$; moreover, according to
SR, its properties with respect to $S$ are the same as in the
preceding case, when instead of accelerating the copy frame, the
object has been accelerated.

Now in this frame-acceleration procedure the problem arises of why
different properties of the object are observed in the $S$ frame, as
compared with its rest properties, apparently without action of any
forces on the object. Fe\u{\i}nberg [10] asked the simple question:
why does the action on the measuring system of the rods and clocks
cause a contraction of the measured rod?.\footnote [9] {Note that
the question is based on the presupposition that just the action on
the measuring rods and clocks of the copy frame is the cause of a
contraction of the measured rod. By the way, in references [10] a
poor translation of Russian originals is found at some places.} The
author stated that the answer is `almost trivial: clearly, if the
measuring instruments are changed somehow under the action of
forces, then the result of the measurement may be changed.' However,
after an explanation that I found somewhat obscure, Fe\u{\i}nberg
stressed that `one may naturally still wonder why a symmetric result
is obtained when there is such an enormous asymmetry in the
transition to the final state of motion with the same relative
velocity.'

The preceding considerations brings us naturally to what is perhaps
the key question of SR. When a single physical object is observed by
two different inertial observers, where do the differences between
their observations come from? Particularly, when an object which is
at rest relative to $S'$ and thus in uniform motion relative to $S$
is observed from the two frames, why the results of observations
differ? (Note that the question has nothing to do with the object's
history either in $S'$ or $S$, the history may be unknown to us.
Note also that the object considered need not be free nor
connected.)

Miller in [2] argued that the differences among observations of
different inertial observers `are due to to the differences in their
respective measuring instruments [the measuring rods and clocks] and
will be referred to as perspectival effects.' Now the problem arises
of where do the differences among their observations come from when
the measuring instruments themselves are observed (say, when a
measuring rod at rest in $S'$ is observed from $S$). The author
explained that `these perspectival effects ultimately have a
dynamical origin because the properties of measuring instruments are
determined by the forces that keep them in equilibrium in their
respective frames.'

At first sight, Miller's explanation appears to be convincing.
Indeed, properties of measuring instruments are determined by forces
that keep them in equilibrium in their respective frames. Since the
forces are generally velocity dependent, equilibrium conditions are
velocity dependent too. (Note that this applies in any inertial
frame.) It follows that equilibrium properties (the length of the
measuring rod and the rate of the clock) are frame dependent, and
thus a change of reference frame involves differences among the
observed properties of the measuring rod and clock (`perspectival
changes').\footnote [10] {This explanation differs from that given
in [2]. Namely, Miller argued, following Fe\u{\i}nberg [10], that
`when the measuring rods and clocks are moved between inertial
observers, they suffer dynamical changes. When the observers use
their dynamically altered rods and clocks to make measurements, it
is not surprising that their results differ and that they differ by
the same factors that are involved in the dynamical changes.' Note
however that measuring instruments need not be transferred between
frames; as was pointed out above, they could be constructed in each
frame `from scratch'. Moreover, even when measuring instruments are
transferred between inertial frames, there always remains the
problem of why the results of observations {\it of the measuring
instruments} by two observers differ (either in initial or final
states). Also, the measuring instruments are {\it not} altered as
observed in their respective rest frames.}

While Fe\u{\i}nberg and Miller advocate a force interpretation of
the so-called kinematical effects of SR, a common thread in their
discussions is that there is no change in the object being observed;
the differences among observations of different inertial observers
are due to the differences in their respective measuring instruments
and the latter are {\it ultimately} due to a change in perspective
(since equilibrium of forces changes in character with a change of
velocity). Thus a change of reference frame is all that matters,
nothing at all has happened to the object being observed.

It seems however that there are some confusing points in the
authors' arguments. First, each inertial observer possesses his or
her own set of measuring instruments which are perfectly identical
to one another.\footnote [11] {This is the content of Born's
`principle of the physical identity of the units of measure'([17],
cf also [13], footnote 12).} A measuring rod at rest in $S$ is in
all respects identical to a measuring rod {\it of the same
construction} at rest in $S'$ under identical physical conditions;
the rods embody the same length in their respective rest frames.
Therefore it is somewhat perplexing to explain the differences
between the observations of the $S$- and $S'$-observer in terms of
the differences in their respective measuring instruments. Deducing,
e. g., the observations of the $S'$-observer through the
corresponding observations of the $S$-observer may lead to
misunderstandings, due to the fact that the relativity of
simultaneity may remain hidden in such deductions. (A metre stick at
rest in $S'$ and parallel to the $x'$-axis is observed in $S$ to
have the length $\sqrt{1 - v^2/c^2}$m but this does not mean that
this reduced length represents a unit of length in $S'$.) Second, it
is rather strange that different {\it dynamical} phenomena in a
physical object (for example, different equilibrium configurations
in a measuring rod\footnote [12] {Miller gave instructive and simple
enough models of a measuring rod and clock in the framework of the
Maxwell-Lorentz theory of electromagnetism that show clearly that
the structure of the measuring instruments is velocity dependent
([2], cf also [18]). Note that Miler's measuring rod is modeled as a
system of point charges which has only one equilibrium configuration
in its rest frame, and thus only one rest frame length. However, a
real connected standard of length may have various equilibrium
configurations in its rest frame and thus various rest frame
lengths. }) are observed as the result of a mere change of reference
frame, apparently without exerting any action upon the object. Since
the $S$ and $S'$ frames (including their respective sets of
measuring instruments) are perfectly equivalent, it seems natural to
look for the root of the differences between the observations in
terms of a change in the object being observed i. e. as the result
of the action of certain forces on it.

\medskip

\noindent {\bf 4. Length contraction puzzle solved? }

\medskip

\noindent Consider a physical object at rest in $S'$. The object
need not be free nor connected. For example, it may consist of two
unconnected stationary material points lying on the $x'$-axis. If
the object is connected, assume that it is in a persistent state.
Let there be another reference frame with its own set of measuring
instruments, perfect copies of $S'$ and its instruments, all at rest
relative to $S'$. Now, following Fe\u{\i}nberg's procedure described
above, accelerate the copy frame together with its instruments in a
rest properties-preserving way with respect to $S'$ in the direction
of the decreasing $x'$ until reaching the steady speed $v$ in all of
its parts (after all transient effects have died away). Thus the
accelerating copy frame eventually becomes our inertial frame $S$
(perhaps after re-synchronizing its clocks, if necessary). Assume
that during the acceleration and after that no action was exerted
upon the physical object being considered from the point of view of
the $S'$-observer. What are the final properties of the object from
the point of view of an observer attached to the copy frame (`the
$C$-observer')?

Construction of the reference frame of an accelerated observer in SR
is somewhat tricky even in the simple case of an observer with a
constant rest acceleration [19]. It seems however that main
conclusions could be reached without entering complex analyses.
Initially, the object was at rest with respect to the inertial copy
frame and its $C$-observer. Then the object was accelerated with
respect to the $C$-observer in the direction of the increasing $x$.
Finally, the object in a persistent state is in uniform motion at
the velocity $\pmb v = v\pmb {\hat{x}}$ with respect to the again
inertial copy frame (now the $S$ frame). Moreover, the object has
{\it a fortiori} the same rest-properties in its final state as it
had in its initial state. (This information could be communicated to
the $C$-observer by a radio transmission.) Since the $C$-observer
finds no differences between the initial and final inertial copy
frame (coinciding with $S'$ and $S$, respectively) he or she rightly
infers that the final properties of the object being considered are
exactly the same as if its complete history developed in the
inertial frame $S$, i. e. as if the object was accelerated with
respect to $S$ starting from rest under the action of certain forces
in a rest properties-preserving way.\footnote [13] {Franklin [1]
recently analysed the case of two unconnected material points that
move relative to an inertial frame with constant rest accelerations
starting from rest in a restlength-preserving way. Some weak points
of [1] are pointed out in [20].}

What actually happened to the copy frame in between is irrelevant
for the final properties of the object. Assume, e. g., that the
$C$-observer had fallen into a deep sleep before the acceleration of
the copy frame began and awoke only after all transient effects of
the acceleration have died away. Thus he or she slept away the
intermediate (non-inertial) stages of the copy frame. Assume also
that the copy frame accelerometer was broken all the time. Then the
$C$-observer would be most inclined to ascribe the change in
velocity of the object (and all related changes in its properties)
to the action of some real external forces upon the object rather
than acceleration of the copy frame. On the other hand, if the
$C$-observer is aware that just the copy frame was accelerated
(either the observer was fully awaken or the accelerometer was in
function during the intermediate stages), he or she could explain
the corresponding acceleration of the object as the result of the
action of some (conditionally speaking) inertial forces (or a
temporarily `switching on' of a gravitational field) as the
classical observer could do. (Needless to say, both the inertial
forces and the gravitational field are just convenient vehicles for
describing the experience of the $C$-observer, without a physical
reality.) Thus `the enormous asymmetry' pointed out by Fe\u{\i}nberg
[10] seems to be removed.

\medskip

\noindent {\bf 4. Conclusions }

\medskip

\noindent When a connected physical object in a persistent state is
observed by observers in two different inertial frames, the
differences between their observations are due to changes in
character of the equilibrium of forces which determines the
structure of the object with a change of its velocity, provided that
the velocity change was performed {\it in a rest
properties-preserving way}. This `perspectival change' in the object
being observed has nothing to do with {\it actual} history of the
object in any of the two inertial frames. However, the perspectival
change may be given a natural dynamical interpretation in terms of a
change in the object under the action of certain forces, either with
respect to {\it any one} inertial frame or with respect to the frame
of an accelerated observer. The last statement applies also to a
system consisting of two or more unconnected material points at
permanent rest relative to an inertial frame. Thus, a change of
reference frame (`a change in perspective') may be understood as
involving a change in the object being observed. The different
inertial observers do have a dynamical explanation of the
differences between their observations in terms of a change in the
object.

It is irrelevant whether two different states of motion of the
object are observed from two different inertial frames,
respectively, or from only one inertial frame, if in the latter case
the two states of motion are related by a rest properties-preserving
acceleration. The results of observations of the object in the two
states of motion are identical in both cases.

\newpage

\noindent {\bf References}

\medskip

\noindent [1] Franklin J 2010 Lorentz contraction, Bell's spaceships
and rigid body motion in special relativity {\it Eur. J. Phys.} {\bf
31} 291-8

\noindent [2] Miller D J 2010 A constructive approach to the special
theory of relativity {\it Am. J. Phys.} {\bf 78} 633-638 (2010)

\noindent [3] Rindler W 1991 {\it Introduction to Special
Relativity} 2nd edn (Oxford: Clarendon) pp 25, 33-6

\noindent [4] Einstein A 1905 Zur Elektrodynamik bewegter K\"
{o}rper {\it Ann. Phys., Lpz.} {\bf 17} 891-921

\noindent [5] Einstein A 1907  Bemerkungen zu der Notiz von Hrn.
Paul Ehrenfest `Die Translation deformierbarer Elektronen und der
Fl\"{a}chensatz,'"  {\it Ann. Phys., Lpz.} {\bf 23} 206-8 (English
translation in Einstein A 1989 {\it The Collected Papers of Albert
Einstein, Vol 2: The Swiss Years: Writings, 1900-1909 (English
Translation Supplement)} (Princeton, NJ: Princeton UP) (translated
by A Beck))

\noindent [6] Pauli W 1958 {\it Theory of Relativity} (London:
Pergamon) (reprinted 1981 transl. G Field (New York: Dover))

\noindent [7] Swann W F G 1941 Relativity, the Fitzgerald-Lorentz
contraction and quantum theory {\it Rev. Mod. Phys.} {\bf 13}
197-202

\noindent [8] J\'{a}nossy L 1971 {\it Theory of Relativity Based on
Physical Reality} (Budapest: Akad\'{e}miai Kiad\'{o})

\noindent [9] Bell J S 1976 How to teach special relativity {\it
Prog. Sci. Cult.} {\bf 1} (2) 1-13, reprinted in

Bell J S 1987 {\it Speakable and Unspeakable in Quantum Mechanics}
(Cambridge: Cambridge UP) pp 67-80

\noindent [10] Fe\u{\i}nberg E L 1975 Can the relativistic change in
the scales of length and time be considered the result of the action
of certain forces? {\it Sov. Phys.Usp.} {\bf 18} 624-35

Fe\u{\i}nberg E L 1997 Special theory of relativity: how good-faith
delusions come about {\it Sov. Phys.Usp.} {\bf 40} 433-5

\noindent [11] Dieks D 1984 The `reality' of the Lorentz contraction
{\it Zeitschrift f\"{u}r allgemeine Wissenschaftstheorie} {\bf 15}
330-42

\noindent [12] Brown H R 2005 a Einstein's misgivings about his 1905
formulation of special relativity {\it Eur. J. Phys.} {\bf 26}
S85-S90

Brown H R 2005 b {\it Physical Relativity: Space-time Structure from
a Dynamical Perspective} (Oxford: Clarendon)

\noindent [13] Red\v zi\' c D V 2008 Towards disentangling the
meaning of relativistic length contraction {\it Eur. J. Phys.} {\bf
29} 191-201

\noindent [14] Rosser W G V 1964 {\it An Introduction to the Theory
of Relativity} (London: Butterworths)

\noindent [15] Schwartz H M 1977 Einstein's comprehensive 1907 essay
on relativity, part II {\it Am. J. Phys.} {\bf 45} 811-7

\noindent [16] Jefimenko O D 1996 Derivation of relativistic force
transformation equations from Lorentz force law {\it Am. J. Phys.}
{\bf 64} 618-620

\noindent [17] Born M 1965 {\it Einstein's Theory of Relativity}
(New York: Dover) (revised edition prepared in collaboration with
G\"{u}nther Leibfried and Walter Biem)

\noindent [18] Sorensen R A 1995 Lorentz contraction: A real change
of shape {\it Am. J. Phys.} {\bf 63} 413-5

\noindent [19] Semay C 2006 Observer with a constant proper
acceleration  {\it Eur. J. Phys.} {\bf 27} 1157-67

\noindent [20] Red\v zi\' c D V 2010 Relativistic length agony
continued arXiv: 1005.4623

\end {document}